\documentclass[aps,pre,amsmath,amsfonts,amssymb,superscriptaddress,preprint]{revtex4}
\usepackage{float}
\usepackage{mathtools}
\usepackage{calrsfs}
\usepackage{times}
\usepackage{latexsym}
\usepackage{graphicx}
\usepackage{amsmath,amssymb}
\usepackage{verbatim,times,bbm}
\usepackage{placeins}
\usepackage{color}
\usepackage[english]{babel}
\usepackage[T1]{fontenc}
\usepackage{graphicx}
\usepackage{amsmath, amsthm, amssymb}
\usepackage{epstopdf}
\usepackage{subfigure}
\usepackage{float}
\usepackage{appendix}
\usepackage{pdfpages}
\usepackage{placeins}
\usepackage{bm}
\usepackage{tikz}
\usetikzlibrary{positioning,shapes}

\newcommand{\affA}{Aix Marseille Univ, Université de Toulon, CNRS, CPT, Marseille, France}
\newcommand{\affB}{CNRS Centre de Physique Th\'eorique UMR7332,
13288 Marseille, France}
\newcommand{\affC}{School of Science and Technology, University of Camerino, I-62032 Camerino, Italy}
\newcommand{\affD}{INFN Sezione di Perugia, I-06123 Perugia, Italy}
\newcommand{\affE}{ QSTAR and INO-CNR, largo Enrico Fermi 2,
             I-50125 Firenze, Italy}

\begin{document}
 \title{Exciting out-of-equilibrium states in macromolecules through light pumping}\label{sec-intro}
 
\author{Elham Faraji}
\email{elham.faraji@unicam.it}
\affiliation{\affC}\affiliation{\affA}\affiliation{\affB}

\author{Roberto Franzosi}
\email{roberto.franzosi@ino.it}
\affiliation{\affE}

\author{Stefano Mancini}
\email{stefano.mancini@unicam.it}
\affiliation{\affC}\affiliation{\affD} 
 
\author{Marco Pettini}
\email{pettini@cpt.univ-mrs.fr}
\affiliation{\affA}\affiliation{\affB}

\date{\today}

\begin{abstract}
In the present paper we address the problem of the energy downconversion of the light absorbed by a protein into its internal vibrational modes. We consider the case in which the light receptors are fluorophores either naturally co-expressed with the protein or artificially covalently bound to some of its amino acids.  In a recent work [Phys. Rev. X 8, 031061 (2018)], it has been experimentally found that by shining a laser light on the fluorophores attached to a protein the energy fed to it can be channeled into the normal mode of lowest frequency of vibration thus making the subunits of the protein coherently oscillate. Even if the phonon condensation phenomenon has been theoretically explained, the first step - the energy transfer from electronic excitation into phonon excitation -  has been left open. The present work is aimed at filling this gap.

\end{abstract}
\maketitle
\section{Introduction}
 The activation of out-of-equilibrium collective intramolecular vibrations of a model protein has been recently reported in Ref. \cite{prx}.  This phenomenon has been induced by light pumping, realised by shining a laser light on an aqueous solution of BSA (Bovine Serum Albumin) protein molecules each one carrying a few fluorophores covalently attached to their Lysine residues. The fluorophores were excited with a blue light at $4880 \mathring{A}$ and then they re-emitted a broadband fluorescence radiation peaked at $5190 \mathring{A}$, thus the difference between the absorbed and re-emitted photon energies resulted in a concentration of an average energy of $0.19$ eV at the fluorophores sites  which thus became "hot points" on each protein. A continuous energy supply of this kind was experimentally found effective to excite the vibrational modes of the proteins and, with an energy supply rate exceeding a suitable threshold, this eventually led to a phonon condensation phenomenon into the lowest vibrational frequency.
The relevance of this out-of-equilibrium collective molecular vibrations consists in the possibility of activating long-range electrodynamic interactions between bio-macromolecules \cite{preto}. The reason is that, at thermal equilibrium, a macromolecule vibrates incoherently with a broad spectrum of modes, whereas the action of an external source of energy promoting a phenomenon of phonon condensation can induce the coherent motion of the molecular subunits, so that, the resulting collective vibration can bring about a large oscillating dipole moment. Under this condition long-range and resonant (thus selective) electrodynamic forces can be activated. In turn, these electrodynamic forces could help explaining the astonishing efficiency of the impressively complex biochemical machinery at work in living cells \cite{bonetta}, where the different actors (proteins, DNA and RNA) find their cognate partners and targets in the right place, at the right time and in the right sequence in an overcrowded environment (the cytosol). Electrodynamic resonant/selective forces are the only possible one to act at a long distance, all the others (chemical bonds, Van der Waals and electrostatic forces) are in fact either intrinsically acting at very short distances, or are screened by the freely moving small ions in the cytosol.  Actually, this is a longstanding theoretical scenario \cite{Frohlich1,Frohlich2,Frohlich3} which, for several reasons, has been discarded. However, the upgrade of Fr\"ohlich's theoretical proposition in \cite{prx,preto} and the experimental outcomes reported in \cite{prx}, represent a first crucial leap forward to ascertain whether the above mentioned hypotheses can be given experimental confirmation or refutation that can be attempted with the nowadays available technology \cite{pre2,pre3}. 

Of course we are faced with the problem of understanding what might replace the laser action in living cells. There are several possible candidates to play the role of external energy suppliers, for instance, the hydrolysis of Adenosine Triphosphate (ATP) releases a highly energetic phosphate group, redox reactions and mitochondria produce weak UV photons that might excite Tryptophan and Tyrosine amino acids \cite{philip,craddock} in proteins, as well nucleotides of DNA and RNA. Also an anisotropic momentum transfer operated by water molecules or ions could make the job \cite{zaragoza}. 
In either cases of metabolically generated photons or of ion collisions (phosphate stemming from ATP hydrolysis or other) we can assume that the external energy input for a biomolecule occurs through the generation of "hot points", as in the case of light activated fluorophores, and mediated by either radiative or collisional electronic excitation. 
In what follows, we aim at better understanding, qualitatively and quantitatively, how part of the photon energy received through electronic excitation of the fluorophores attached to a biomolecule (protein) is converted into vibrational energy of the chain of subunits (amino acids) composing it. As we shall see, it is found that only a fraction of the initially available electron energy is released to the phonons of a biomolecule. The quantitative estimate of this energy transfer process is very important for a better assessment of the physical conditions which are necessary to activate the intramolecular collective vibrations.

\section{Definition of the model}
In Ref.\cite{prx} the external source of energy driving the phonon condensation was modeled (by one of us among the others) as high temperature heat bath. This was done to reformulate in a classical framework the Wu-Austin \cite{wuaustin} quantum model leading to the original Fr\"ohlich rate equations of Ref.\cite{Frohlich1}. We now aim at refining this part of the model in view of a better understanding of the basic excitation mechanism that can bring a macromolecule out of thermal equilibrium.

In both cases of photo-excitation and, presumably, of ionic collisions, the excitation mechanism is supposed to be mediated by the molecular electron cloud.
Therefore, the model describing the phenomenon that we want to investigate is borrowed from the standard Davydov and Holstein-Fr\"ohlich models \cite{standard,froehpolaron,holstein} to account for electron-phonon interaction. Hence, the following energy operator is assumed    
   \begin{eqnarray}\label{H}
   \hat{H} =\hat{H}_{el}+\hat{H}_{ph}+\hat{H}_{int},
   \end{eqnarray}
where the first term $ \hat{H}_{el} $ is the electron energy operator   
   \begin{equation}\label{Hex}
   \hat{H}_{el}=\sum_{n=1}^{N}
   \Big[ E_{0}\hat{B}_{n}^{\dag} \hat{B}_{n}+\epsilon \langle\hat{B}_{n}^{\dag } \hat{B}_{n}\rangle \hat{B}_{n}^{\dag } \hat{B}_{n}+J(\hat{B}_{n}^{\dag} \hat{B}_{n+1}+\hat{B}_{n}^{\dag} \hat{B}_{n-1})\Big],
      \end{equation}
with $\hat{B}_n$ and $\hat{B}^{\dag}_n$ the annihilation and creation operators for the electron at any site $n$ $(n=1,2,,...,N)$ which labels the amino acid along the protein. The term $E_{0}\hat{B}_{n}^{\dag} \hat{B}_{n}$ accounts for the initial "bare" electron energy distributed on several lattice sites according to initial shape of the electron wavefunction.  The constant $J$ is the nearest neighbour coupling energy of the hopping electron across two neighbouring amino acids. In this model we have considered only a longitudinal chain of amino acids. 
The electron moving from the excited fluorophore interacts on its way with almost free electrons in each amino acid, and it may just make a disturbance which will allow a next electron to continue on the trip. It would be then more a disturbance traveling than a single electron, but the net effect will be the same of a traveling electron.
Thus the term $\epsilon \langle\hat{B}_{n}^{\dag } \hat{B}_{n}\rangle \hat{B}_{n}^{\dag } \hat{B}_{n} $ has been introduced to take into account non-linear effects due to the interaction between the electron in motion along the chain and the electrons of the substrate of amino acids. In particular, the term takes into account effects related to the Coulombic repulsion between the traveling electron and the charges localized on the amino acids. The averaging is intended as the expectation value of $\hat{B}_{n}^{\dag } \hat{B}_{n}$ on the dynamically evolving state of the system.

 The second term $ \hat{H}_{ph} $ in \eqref{H} is the phonon energy operator
 \begin{equation}\label{Hph}
 \hat{H}_{ph}=\frac{1}{2}\sum_{n}\Big[\frac{\hat{p}_{n}^2}{M}+\Omega(\hat{u}_{n+1}-\hat{u}_{n})^2+{1\over 2}\mu (\hat{u}_{n+1}-\hat{u}_{n})^{4}\Big],
   \end{equation} 
   where $\hat{p}_{n}$ and $\hat{u}_{n}$ are momentum and position operators for longitudinal displacements of amino acids at site $n$, respectively. Furthermore, $M$ and $\Omega$ are average values of the mass of the amino acids of a protein and of the spring constants of two neighbouring amino acids, respectively. The quartic term is a correction stemming from the power series which gives the harmonic term at the lowest order expansion around the minimum of interparticle interaction potential (typically nonlinear, as is the case, for example, of the Van der Waals potential). This term is responsible for phonon-phonon interaction, absent in the harmonic approximation.
   
 Finally,  the third term $ \hat{H}_{int} $ in \eqref{H} is the electron-phonon interaction operator 
 \begin{equation}\label{Hin}
 \hat{H}_{int}=\sum_{n}\chi(\hat{u}_{n+1}-\hat{u}_{n})\hat{B}_{n}^{\dag}\hat{B}_{n},
   \end{equation} 
   where $\chi$ is the energy coupling parameter.
   
   
  \section{Derivation of the dynamical equations with TDVP}
 In order to derive from the model Hamiltonian \eqref{H} the corresponding dynamical equations,  we make a simplifying ansatz about the state vectors by
 assuming  the following factorization
  \begin{equation}\label{psi}
|\psi\rangle=|\Psi\rangle|\Phi\rangle 
   \end{equation} 
in which $|\Psi\rangle$ describes an electron given a single quantum excitation and supposed to be free to propagate along the chain of $N$ amino acids composing a protein 
     \begin{equation}\label{exiton}
    |\Psi(t)\rangle =\sum_{n}C_{n}(t)\hat{B}_{n}^{\dag}|0\rangle_{el},
     \end{equation}
where $|0\rangle_{el}$ is the vacuum state of the Amide-I oscillators, and
             \begin{equation}\label{phononavr}
|\Phi(t)\rangle=e^{-{i \over \hbar}\sum[\beta_{n}(t)\hat{p}_{n}-\pi_{n}(t)\hat{u}_{n}]}|0\rangle_{ph}.
   \end{equation} 
We then set
           \begin{eqnarray}\label{22}
\langle\Phi|\hat{u}_{n}|\Phi\rangle &=&\beta_{n}(t),\nonumber\\
\langle\Phi|\hat{p}_{n}|\Phi\rangle &=&\pi_{n}(t),
   \end{eqnarray} 
where $\beta_{n}(t)$ and $\pi_{n}(t)$ are the average values of the longitudinal displacement and momentum of an amino acid, respectively.

 To derive dynamical equation we now resort to the time-dependent variational principle (TDVP) in quantum mechanics. TDVP is a formulation of the time-dependent
Schr\"{o}dinger equation through variation of an action functional. The Schr\"{o}dinger equation is obtained by requiring that the action functional be stationary under free variation of the time-dependent state. According to this principle, we define a new wave function $|\phi\rangle$ in terms of $|\psi\rangle$ in Eq. (\ref{psi}) as
 \begin{eqnarray}\label{psitdvp}
|\phi(t)\rangle=e^{iS(t)/\hbar}|\psi(t)\rangle,
   \end{eqnarray} 
  where $S(t)$ is a time-dependent phase factor $(S(t)\in \mathbb{R})$, which will be determined in a self-consistent manner and the normalization condition is $\langle \phi|\phi\rangle=1$. The wave function  $|\phi\rangle$ satisfies the Schr\"{o}dinger equation 
        \begin{eqnarray}
i\hbar\langle\phi(t) | \partial_t|\phi(t)\rangle=\langle\phi(t) |\hat{H}|\phi(t)\rangle,
   \end{eqnarray} 
   which according to Eq. (\ref{psitdvp}) becomes
           \begin{eqnarray}
-\dot{S}(t)+i\hbar\langle\psi(t) |\partial_t|\psi(t)\rangle=\langle\psi(t) |\hat{H}|\psi(t)\rangle.
   \end{eqnarray} 
Integrating, we obtain
           \begin{eqnarray}
S(t)=\int_{0}^{t} \Big[i\hbar\langle\psi(t) |\partial_t|\psi(t)\rangle-\langle\psi(t) |\hat{H}|\psi(t)\rangle\Big]dt.
   \end{eqnarray} 
We can now derive the equations of motion by requiring that the action with the Lagrangian 
\begin{eqnarray}\label{LL}
L=i\hbar \langle\psi(t) |\partial_t|\psi(t)\rangle-\langle\psi(t) |\hat{H}|\psi(t)\rangle\ ,
   \end{eqnarray} 
   to be stationary
      \begin{eqnarray}\label{stationary}
\delta S(t)=\delta \int L dt=0.   
\end{eqnarray} 
From Eqs. (\ref{psi}), (\ref{exiton}), and (\ref{phononavr}) we write
        \begin{eqnarray}
\partial_{t}|\psi\rangle=\left(\partial_{t}|\Psi\rangle\right)|\Phi\rangle
+|\Psi\rangle \left(\partial_{t}|\Phi\rangle\right),
   \end{eqnarray}
   and then arrive at
\begin{equation}\label{111}
\langle \psi|\partial_{t}|\psi\rangle=\sum_{n}\left[\dot{C}_{n}(t)C_{n}^{*}(t)+{i\over 2\hbar}\Big(\dot{\pi}_{n}(t)\beta_{n}(t)-\pi_{n}(t)\dot{\beta}_{n}(t)\Big)\right].
   \end{equation}
  Thus the Lagrangian \eqref{LL} becomes
    \begin{eqnarray}
L=\sum_{n}\left\{i\hbar \dot{C}_{n}(t)C_{n}^{*}(t)+{1\over2}\Big(\pi_{n}(t)\dot{\beta}_{n}(t)-\dot{\pi}_{n}(t)\beta_{n}(t)\Big)-H(C_{n},C_{n}^*,\beta_{n},\pi_{n})\right\},
   \end{eqnarray}
   where
       \begin{eqnarray}
H(C_{n},C_{n}^*,\beta_{n},\pi_{n})= \langle\psi(t) |\hat{H}|\psi(t)\rangle.
   \end{eqnarray}
Imposing the condition \eqref{stationary}, we get 
\begin{eqnarray}
\delta S(t)&=\sum_{n}\Big\{ i\hbar\Big(-\dot{C}_{n}^{*}(t)\delta C_{n}(t)+\dot{C}_{n}(t)\delta C_{n}^{*}(t)\Big)+\dot{\beta}_{n}(t)\delta \pi_{n}(t)-\dot{\pi}_{n}(t)\delta \beta_{n}(t)\nonumber\\
&-(\partial_{C_{n}}H)\delta C_{n}-(\partial_{C_{n}^{*}}H)\delta C_{n}^{*}-(\partial _{\beta_{n}}H)\delta \beta_{n}-(\partial_{\pi_{n}}H)\delta \pi_{n}\Big\}=0,
   \end{eqnarray}
    from which it results
           \begin{eqnarray}\label{hc}
i\hbar\dot{C}_{n}&=&\partial_{C^*_{n}} H\nonumber\\
\dot{\beta}_{n}&=&\partial _{\pi_{n}} H\nonumber\\
\dot{\pi}_{n}&=&-\partial _{\beta_{n}} H \ .
   \end{eqnarray}
The expectation value of the Hamiltonian is
      \begin{eqnarray}\label{Hnew}
\langle \psi|\hat{H}|\psi \rangle&=\sum_{n}\Big[E_{0}|C_{n}|^{2}+\epsilon|C_{n}|^{4}+ J(C_{n}^{*}C_{n+1}+C_{n+1}^{*}C_{n})\nonumber\\
&+ {1\over 2}\Big({\pi_{n}^2\over M}+\Omega (\beta_{n+1}-\beta_{n})^2+{1\over 2}\mu (\beta_{n+1}-\beta_{n})^4\Big),\nonumber\\
&+\chi(\beta_{n+1}-\beta_{n})|C_{n}|^2\Big].
   \end{eqnarray}
   So, from Eq. (\ref{Hnew}) we have
 \begin{eqnarray}\label{dynamicaleqs}
  i\hbar\dot{C}_{n}&=&\Big(E_{0}+2\epsilon|C_{n}|^{2}+\chi(\beta_{n+1}-\beta_{n})\Big)C_{n}
 +J(C_{n+1}+C_{n-1}),\nonumber\\
 M\ddot{\beta}_{n}&=&\Omega(\beta_{n+1}-2\beta_{n}+\beta_{n-1})+\chi \Big(|C_{n}|^2-|C_{n-1}|^2\Big)\nonumber\\
&+&\mu \Big((\beta_{n+1}-\beta_{n})^{3}-(\beta_{n}-\beta_{n-1})^{3}\Big).
    \end{eqnarray}

  \section{Definition of the physical parameters for numerical simulations}  
 {{Let us see how to make a physically reasonable choice of the coupling parameters entering the Hamiltonian. We borrow from  Ref.\cite{Cosic,pseudopot} the estimates of the interaction energy between an electron and each of all the 20 amino acids. The average value of these interaction energies is $\langle\Delta E\rangle =0.74$ eV with a dispersion $\sigma_E=0.47$ eV. As a first rough picture of an electron hopping across the sequence of amino acids constituting a protein we can consider the electron of energy $E_0$ moving in a periodic sequence of square potential barriers of height $V_0=0.74$ eV and of width $a = 4.5 \mathring{A}$, the average distance between two nearest neighboring amino acids \cite{standard}. We can then weigh the electron hopping operators between neighbouring sites with the probability $P(n\to n\pm 1)$ of tunnelling from one potential well to the nearest ones. This is achieved by computing the transmission coefficient
 \begin{equation}
 T= \left[ 1 + \frac{ V_0^2 \sinh^2\beta a}{4 E_0(V_0 - E_0)}\right]^{-1}
 \end{equation}
where $\beta =[2 m_e(V_0 - E_0)/\hbar^2]^{1/2}$.  Moreover, the coefficient of the electron hopping term in the Hamiltonian has to be a characteristic energy scale of the process, thus a natural choice is to set $J \propto\langle\Delta E\rangle T$, then, assuming that an electron is initially excited at any given point of the chain of amino acids and that it has the same probability of moving to the left or to the right, we add a factor $1/2$ so that finally we have   $J =\frac{1}{2}\langle\Delta E\rangle T$. Now, assuming $E_0=0.19$ eV as initial value of the electron energy, we find
$J=0.0585$ eV, whereas assuming that only a fraction $\eta\in[0,1]$ of the maximum available energy is kept by the electron, for example for $\eta=0.5$, we find $J = 0.031$ eV. For what concerns the electron-phonon coupling constant $\chi$, we make a rough estimate of its value as $\chi =\Delta E/\Delta x = \sigma_E/\Delta x = \sigma_E/a = 0.47 eV/4.5\mathring{A}\simeq 100$ pN.
    
In what follows, in dimensionless units, we have $\chi' =0.81$, and $J'=5$ with $\eta =0.5$, while $J'=9$ with $\eta =1$.

By rescaling time and lengths as $t=\omega^{-1} \tau$ and $\beta_{n}=L b_{n}$, respectively, where $L=\sqrt{\hbar \omega^{-1}M^{-1}}$, the following dimensionless dynamical equations are obtained
\begin{eqnarray}\label{dimdynamics}
  i\frac{d{C}_{n}}{d\tau}&=&\Big[\Big(E'+2\epsilon'|C_{n}|^{2}+\chi'(b_{n+1}-b_{n})\Big)C_{n}
 +J'(C_{n+1}+C_{n-1})\Big],\nonumber\\
\frac{d^{2}{b}_{n}}{d\tau^2}&=&\Omega'(b_{n+1}-2 b_{n}+b_{n-1})+\chi'\Big(|C_{n}|^2-|C_{n-1}|^2\Big)\nonumber\\
&+&\mu'\Big[(b_{n+1}-b_{n})^3-(b_{n}-b_{n-1})^3\Big],
   \end{eqnarray}
and the dimensionless expression of the Hamiltonian is
         \begin{align}\label{Hsimpli}
\langle \psi|\hat{H}|\psi \rangle&=\sum_{n}\Big[E'|C_{n}|^{2}+\epsilon'|C_{n}|^{4}+ J'(C_{n}^{*}C_{n+1}+C_{n+1}^{*}C_{n})\nonumber\\
&+ {1\over 2}\Big(\dot{b}^{2}_{n}+ \Omega'(b_{n+1}-b_{n})^2+{1\over 2}\mu'(b_{n+1}-b_{n})^4\Big),\nonumber\\
&+\chi'(b_{n+1}-b_{n})|C_{n}|^2\Big],
   \end{align}
where
   \begin{eqnarray}\label{dimparameter}
   E'&=&{E_{0} \over \hbar \omega};\;\;\;\;\;\;\; \epsilon'={\epsilon \over \hbar \omega};\;\;\;\;\;\;\; J'={J\over \hbar \omega};\;\;\;\;\;\;\; \nonumber\\
    \chi'&=&{\chi \over \sqrt{\hbar M\omega^{3}}};\;\;\;\;\; \Omega'={\Omega\over M \omega^2};\;\;\;\;\;\;\mu'={\mu \hbar \over M^{2}\omega^{3}}\ .
   \end{eqnarray}
   
In order to perform numerical integration of the dynamical equations it is useful to introduce  the variables
           \begin{eqnarray}
q_{n}={C_{n}+C_{n}^{*}\over \sqrt{2}}, \qquad p_{n}={C_{n}-C_{n}^{*}\over i\sqrt{2}}\ ,
   \end{eqnarray}
so that Eqs.(\ref{dimdynamics}) become
\begin{align}
           \dot{q}_{n}&=\Big[E'+{\epsilon' \over 2}(q_{n}^{2}+p_{n}^{2})+\chi'(b_{n+1}-b_{n})\Big]p_{n}+J'(p_{n+1}+p_{n-1}),\label{differential28}\\
\dot{p}_{n}&=-\Big[E'+{\epsilon' \over 2}(q_{n}^{2}+p_{n}^{2})+\chi'(b_{n+1}-b_{n})\Big]q_{n}+J'(q_{n+1}+q_{n-1})\Big], \label{differential29}\\
\ddot{b}_n&=\Omega'(b_{n+1}-2b_{n}+b_{n-1})+{\chi'\over2}   \Big(
(q_{n}^{2}+p_{n}^{2})-(q_{n-1}^{2}+p_{n-1}^{2})\Big)\nonumber\\
&+\mu'\Big[(b_{n+1}-b_{n})^3-(b_{n}-b_{n-1})^3\Big]\label{differentialqp}.
   \end{align}
By denoting with $\mathcal{B}_n[\textbf{b}(t), \textbf{q}(t), \textbf{p}(t)]$ the r.h.s. of  Eq. \eqref{differentialqp} we have
\begin{equation}
b_n(t+\Delta t) = 2 b_n(t) - b_n(t - \Delta t) + (\Delta t)^2 \mathcal{B}_n[\textbf{b}(t), \textbf{q}(t), \textbf{p}(t)]
\end{equation}  
which can be rewritten in the form
\begin{eqnarray}\label{bienne}
\dot{b}_n&=& \pi_n\nonumber\\
\dot{\pi}_n&=& \mathcal{B}_n[\textbf{b}(t), \textbf{q}(t), \textbf{p}(t)]\ .
\end{eqnarray} 
Equations \eqref{differential28} and \eqref{differential29} and the above system have been numerically integrated by combining a finite differences scheme and a leap-frog scheme as follows
   \begin{eqnarray}
q_{n}(t+\Delta t)&=& q_{n}(t)+\Delta t\ \mathcal{Q}_n[\textbf{b}(t), \textbf{q}(t), \textbf{p}(t)],\nonumber\\
p_{n}(t+\Delta t)&=& p_{n}(t)+\Delta t\ \mathcal{P}_n[\textbf{b}(t), \textbf{q}(t), \textbf{p}(t),\nonumber\\
b_{n}(t+\Delta t)&=& b_{n}(t)+\Delta t\ \pi_n(t)\nonumber\\
\pi_{n}(t+\Delta t)&=& \pi_{n}(t)+\Delta t\ \mathcal{B}_n[\textbf{b}(t+\Delta t), \textbf{q}(t+\Delta t), \textbf{p}(t+\Delta t)] .
   \end{eqnarray}
   where $\mathcal{Q}_n[\textbf{b}(t), \textbf{q}(t), \textbf{p}(t)]$ and $\mathcal{P}_n[\textbf{b}(t), \textbf{q}(t), \textbf{p}(t)$ are the r.h.s. of Eqs.\eqref{differential28} and \eqref{differential29}, respectively.
 By using sufficiently small time steps $\Delta t$ the desired precision of energy conservation can be attained.
  
 About the initial conditions, we aim at simulating a physical situation where 
 each photon absorbed by a fluorophore attached to a protein  releases  - in the average - $0.19$ eV of energy to the surrounding electron cloud. This energy is the difference between the energies of the absorbed photon of $4880 \mathring{A}$ and that of the re-emitted one as fluorescent radiation of $5150 \mathring{A}$.  We assume, as already stated above, that the effect of a single photon excitation is to make one electron moving across the protein by tunnelling through a sequence of potential barriers. In the experiments to which we are referring \cite{prx} each protein is labelled with 5-6 fluorochromes, and a laser light is continuously shined on the labelled proteins, therefore what we are after is modelling an elementary process and assuming, in a first approximation, a property of additivity of the same elementary process. In other words, if more than one electron is activated we assume that the resulting physical effect is the sum of a single electron effect.
As a consequence, the electron initial condition is assumed to be described by a wavefunction $C_{n}(t=0)$ centered at the site $n=n_{0} $ at time $t=0$ \cite{standard}:
         \begin{eqnarray}\label{exc}
C_{n}(t=0)&=& {1\over \sqrt{8 \sigma_{0}}}{\rm sech}\Big({n-n_{0}\over 4\sigma_{0} }\Big).
    \end{eqnarray}\\
where $\sigma_{0}=3 \Omega J/\chi^{2}$.\\
Then, coming to the initial conditions of the phonon component of the system, we assume a thermalized macromolecule at room temperature, that is at $T =310 K$. At equilibrium, the energy equipartition theorem for the Hamiltonian (\ref{Hph}) reads
      \begin{eqnarray}
\left\langle p_{n}{\partial H_{ph} \over \partial p_{n}}\right\rangle=\left\langle u_{n}{\partial H_{ph} \over \partial u_{n}}\right\rangle=k_{B}T
    \end{eqnarray}
where $k_{B}$ is the Boltzmann constant.
At thermal equilibrium,  energy is equally shared among all the degrees of freedom and, in particular,  between kinetic and potential energies, therefore at  $t=0$ the velocities and the displacements have been initialized with random values of zero mean and fulfilling the conditions
   \begin{equation}\label{thermequil}
\langle\vert b_{n}(0)\vert\rangle_n=\sqrt{\frac{k_{B}T}{\hbar \omega \Omega'}} ;\;\;\;\;\;\;\;\;\;\;\;\;\;\;\;\;\; \langle\vert \dot{b}_{n}(0)\vert\rangle_n=\sqrt{\frac{k_{B}T}{\hbar \omega}}.
    \end{equation}
expressed in dimensionless form.

In Table \ref{parameters1} the values chosen for the physical parameters are reported. These are: the initial excitation energy $E_{0}$, an average value of the mass $M$ of the amino acids, the dipole-dipole coupling constant $J$, the elasticity constant $\Omega$ used in the numerical studies of Ref. \cite{standard}, and the electron-phonon coupling $\chi$. In Table \ref{parameters1} also the corresponding dimensionless values of the same physical quantities are reported, these are obtained by using (\ref{dimparameter}) and the frequency  $\omega=10^{13} s^{-1}$.\\
\begin{table}
\begin{tabular}{ccccc }
\hline
\hline
Name &      Symbol &	Value & Symbol & Dimensionless value  \\ 
\hline
\hline
Hot-point energy & $E_{0}$ &	0.2 eV & E' & 30\\ 
\hline
 Average mass of amino acids &      M&	1.5 $\times 10^{-25}$ kg &- & - \\ 
\hline
Spring constant &	 $\Omega$ & 18.3 N/m & $\Omega'$ & 1.2\\
\hline
Electron hopping parameter&	J & 0.0658 eV &J'& 10\\
\hline
Electron-phonon coupling  &	$\chi$ & 61-610 pN & $\chi'$ & 0.5-5\\
\hline
Anharmonic parameter  &	$\mu$ & Arbitrary & $\mu'$ & 0-0.5\\
\hline
Nonlinear parameter  &	$\epsilon$ & 0.00658-0.065.8 eV & $\epsilon' $ & 1-10\\
\hline
\end{tabular}
\caption{Values of the parameters used in the numerical simulations. Physical versus dimensionless values are reported.}
\label{parameters1}
\end{table}
\section{Numerical results}
All the numerical computations have been performed using an integration time step  $\Delta t=5 \times 10^{-5}$ 
entailing a very good energy conservation, with typical relative error $\Delta E/E\simeq 10^{-5}$. 
 The length of the chain is $N=500$ rounding the number of amino acids of the protein in \cite{prx}. Figures (\ref{amplitude2}) and (\ref{amplitude30}) show the spatial distribution of the probability $\vert\psi(n,t)\vert^2$ of finding the moving  electron at any site $n$ versus  time for the electron-phonon coupling $\chi=100$ pN and $\chi=366$ pN, respectively. The electron is initially centered around the site $n=250$.  Figure (\ref{amplitude2}) shows that the electron wavefunction quickly spreads over the whole substate of amino acids, a phenomenon somewhat less pronounced in Figure (\ref{amplitude30}) and to some extent counterintuitive since the latter corresponds to a stronger electron-phonon coupling. \\
Figure (\ref{displacement}) shows the time evolution of random initial conditions for the displacements of the underlying chain of masses modelling the chain of amino acids of a protein. The random initial displacements and velocities are generated at thermal equilibrium at 310 K, according to the prescriptions of Eq.\eqref{thermequil}. \\
Figure (\ref{energies}) synoptically displays the energy transfer from the electron to the phonon subsystem. The same figure also shows that the larger $\chi$ the faster this energy transfer, what is physically sound and not necessarily at odds with what reported in Figures (\ref{amplitude2}) and (\ref{amplitude30}) about the electron wavefunction spreading.

As is seen from the plots in Figures (\ref{mu}), the value of the phonon-phonon coupling parameter $\mu'$ does not seem crucial to control the release of the electron energy to the phonons, the process appears to be mainly driven by the electron-phonon coupling constant. In fact, for $\chi=488$ pN the relaxation to the oscillatory state is quick and practically independent of the value of $\mu'$. At the lower value $\chi=61$ pN some differences in the relaxation rate are observed by varying $\mu'$, but even for $\mu' =0$ the energy transfer takes place in both cases of $\chi=61$ pN and $\chi=488$ pN. 

Then we have checked how the phenomenology changes as a consequence of the introduction of the nonlinear coupling in the electron Hamiltonian. 
In Figures (\ref{epsilonhalf}) and (\ref{epsilonfour}) the effects of different values of the parameter $\epsilon$ are reported, again for $\chi=61$ pN and $\chi=488$ pN
respectively. Again for $\chi=488$ pN the electron energy fastly decrease in time, apart from the case of $\epsilon=6.58$ meV where it displays wide oscillations.
At $\chi=61$ pN the electron energy relaxation is slower and for $\epsilon=6.58$ meV it appears to be very slow. \\
Let us remark that a non-vanishing value of $\epsilon$, that is, the presence of the nonlinear coupling term in the electron Hamiltonian, plays a relevant role 
to ensure a more efficient transfer of part of the electron energy to the phonons of the chain of amino acids.

For any chosen set of physical parameters, except possibly for $\epsilon =0$,  the electron always transfers part of its energy to the phonons, and eventually this energy is  equally shared among the phonons. In order to work out the typical time scales of this process we have computed the spectral entropy of the normal modes of the chain of amino acids, that is, of the phonons.
For the harmonic term $H_{h}$ of the dimensionless Hamiltonian (\ref{Hsimpli}) we have
 \begin{align}\label{HarmonicH}
\langle \psi|\hat{H}_{h}|\psi \rangle={1\over 2}\sum_{n=1}^{N}\Big[\dot{b}^{2}_{n}+ \Omega'(b_{n+1}-b_{n})^2\Big],
   \end{align}
and then, by following Ref. \cite{transformation}, the coordinate transformations $Q_m=S_{mn}b_{n}$ and $P_m=S_{mn}\dot{b}_n$,  with 
\begin{equation}
S_{mn}={1 \over \sqrt{N}}\Big[\cos({2\pi \over N}mn)+\sin({2\pi \over N}mn)\Big]\;\;\;\; m,n =1,2,..,N\ ,
\end{equation}
transform the Hamiltonian (\ref{HarmonicH}) into
\begin{equation}
\tilde{H}_h = {1 \over 2}\sum_{m=1}^{N}(P^{2}_{m}+\Omega^{'} \omega^{2}_{m}Q^{2}_{m}),
\end{equation}
where
\begin{equation}
\omega^{2}_{m}=4\sin^{2}(\frac{\pi m}{N}).
\end{equation}
Of course, these oscillators are the normal modes (phonons) of the system. Then a spectral entropy $S(t)$ is defined as
\begin{equation}
S(t)=-\sum_{m=1}^{N}p_{m}(t) \ln p_{m}(t);\;\;\;\;\;\;\;\; p_{m}(t) =\frac{ E_{m}(t)}{E_{T}(t)}
\end{equation}
where $E_{T}(t)=\sum_{m=1}^{N} E_{m}(t)$ and $E_{m}(t) =(P^{2}_{m}+\Omega^{'} \omega^{2}_{m}Q^{2}_{m})/2$, so that the weights $p_{m}(t)$ are normalized. The maximum value of $S(t)$ is attained when all the $p_{m}(t)$ are equal to $1/N$. Thus, at equipartition, when the energy content of each normal mode is the same, entropy attains its maximum, this allows to define a normalized entropy as
\begin{equation}\label{etaeq}
\eta(t)=\frac{S_{max}(t)-S(t)}{S_{max}(t)-S(0)},
\end{equation}
so that when the phonon oscillators are "frozen" it is $S(t)=S(0)$ and consequently $\eta=1$; but at equipartition, when $S(t)=S_{max}(t)$, it is $\eta=0$. 
By following the time decay of  $\eta$, it is thus possible to find out if and on which time scale the energy released by the electron is definitely transferred to the phonons.
In Figure (\ref{eta}) $\eta(t)$ is plotted as a function of time for various values of the coupling constant $\chi$ and keeping fixed the other parameters as in the case reported in Figure (\ref{amplitude2}). It is evident that equipartition of energy is always attained, and the time needed for this to happen is rather weakly dependent on the electron-phonon coupling constant. In fact, the decay time is approximately varying between 0.5 ns and 1 ns (the unit time scale being $10^{-13}$ seconds). Let us remark that the two time scales of the electron energy release to the amino acids and of equipartition of this energy among all the normal modes of the lattice are not equal, and need not to be equal.
    \begin{figure}[h!]
   \centerline{\includegraphics[width=0.95\columnwidth]{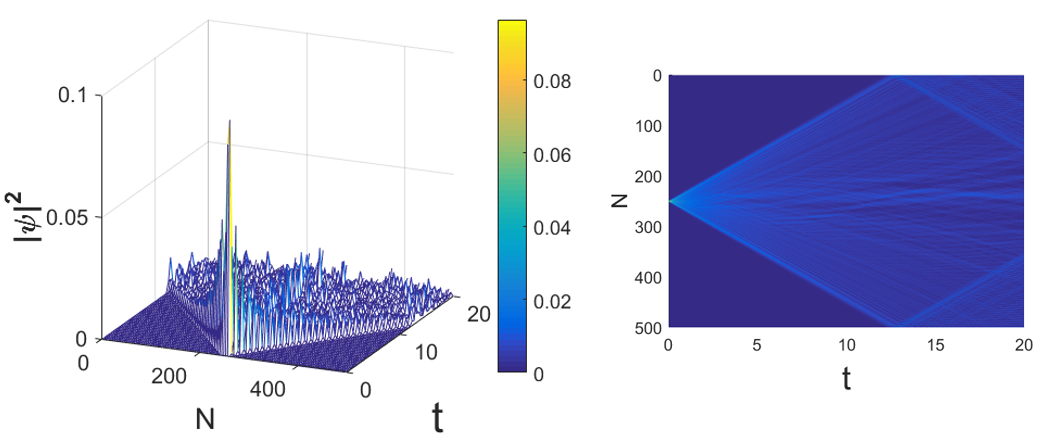}}
   \caption{Evolution of the  probability amplitude of an electron $|\psi(t)|^2$ along the chain of $N=500$ amino acids. Initial conditions: $T=310^{\circ}$K, 
   $E'=30$, $J'=10$, $\epsilon'=5$, $\chi'=0.8$, $\Omega'=1.2$, $\mu'=0.1$, corresponding to $E_0=0.2$ eV, $J=0.0658$ eV, $\epsilon=0.0329$ eV, $\chi=100$ pN, $\Omega=18.3$ N/m, respectively. The right figure is the above view of the left one. Time $t$ is measured in $10^{-13}$s.}
          \label{amplitude2}
   \end{figure}
  
    \begin{figure}[h!]
   \centerline{\includegraphics[width=0.95\columnwidth]{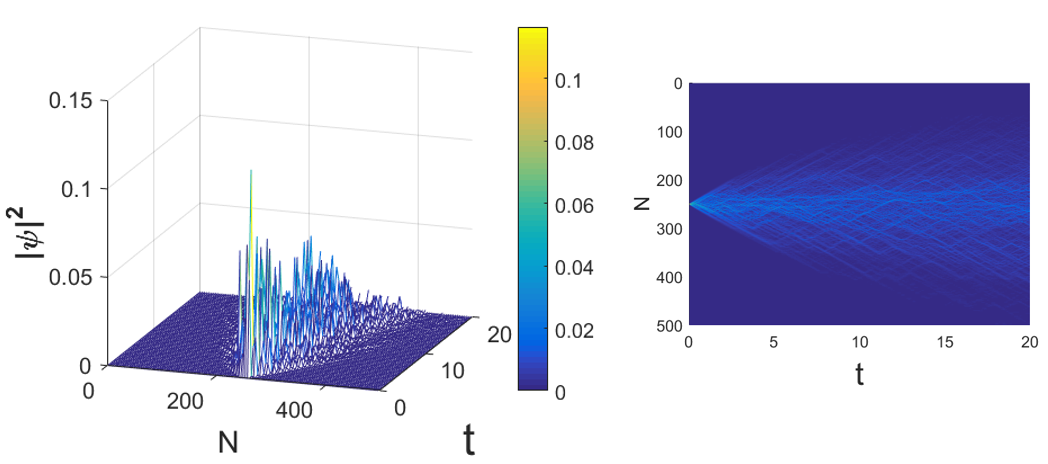}}
   \caption{Evolution of the probability amplitude of an electron $|\psi(t)|^2$ with $N=500$ and $\chi'=3$ ($\chi=366$ pN); the other parameters are the same of Fig. \ref{amplitude2}. Time $t$ is measured in $10^{-13}$s.}
          \label{amplitude30}
   \end{figure}
   
 \begin{figure}[h!]
   \centerline{\includegraphics[width=0.95\columnwidth]{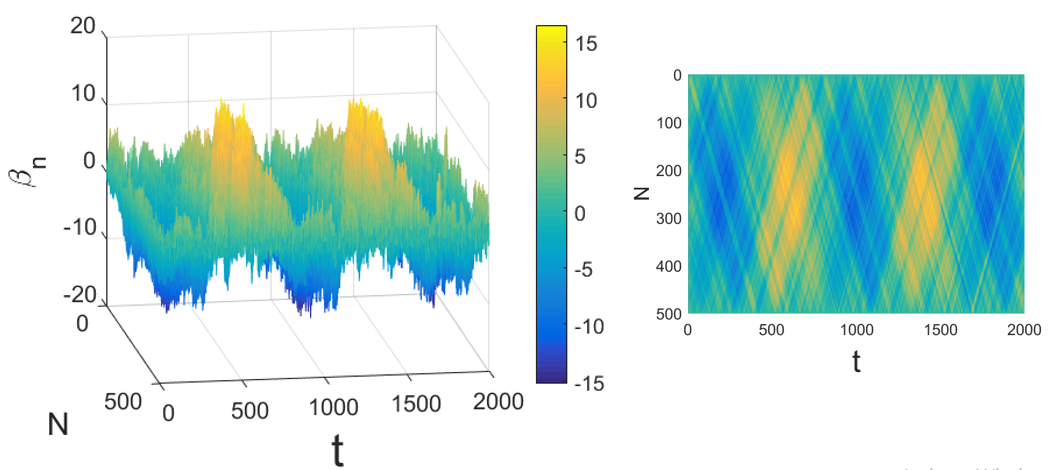}}
   \caption{Time evolution of the average displacements along the chain of $N=500$ amino acids. The parameter values are the same of Fig. \ref{amplitude2}. Time $t$ is measured in $10^{-13}$s.}
          \label{displacement}
   \end{figure}
   
    \begin{figure}[h!]
   \centerline{\includegraphics[width=0.95\columnwidth]{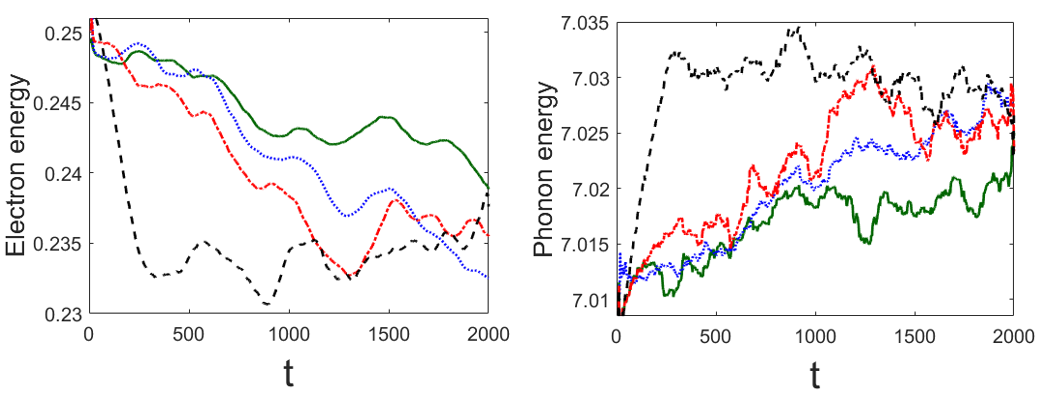}}
   \caption{Energy transfer from the electron to the phonons for $\chi'=0.5$ ($\chi=73.2$ pN) (green solid line), $\chi'=0.8$ ($\chi=100$ pN) (blue dotted line), 
   $\chi'=1$ ($\chi=122$ pN) (red dot-dashed line), and $\chi'=1.5$ ($\chi=183$ pN) (black dashed line); the other parameters are the same of Fig. \ref{amplitude2}. Time $t$ is measured in $10^{-13}$s; electron energy and total phonon energy are given in eV.}
          \label{energies}
   \end{figure} 
       \begin{figure}[h!]
   \centerline{\includegraphics[width=0.95\columnwidth]{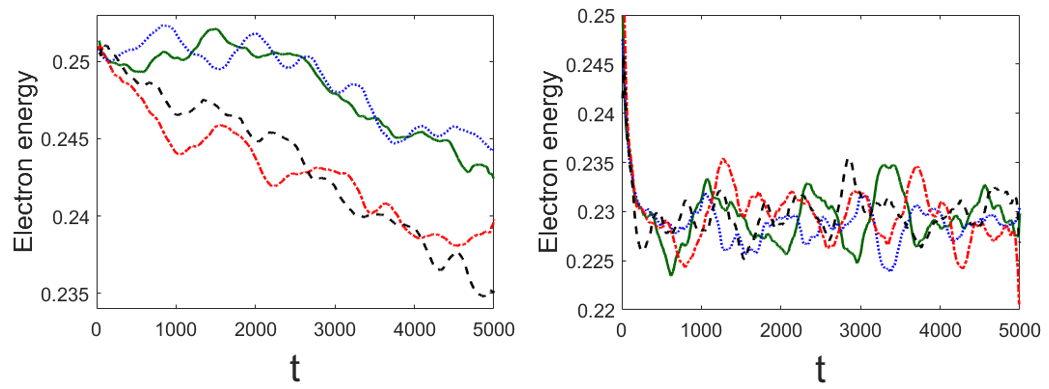}}
   \caption{Decay of the electron energy for $\mu'=0$ (green solid line), $\mu'=0.1$ (blue dotted line), $\mu'=0.3$ (red dot-dashed line), and $\mu'=0.5$ (black dashed line); the other parameters are the same of Fig. \ref{amplitude2}, except for $\chi'=0.5$ ($\chi=61$ pN) (left panel) and $\chi'=4$ ($\chi=488$ pN) (right panel). Time $t$ is measured in $10^{-13}$s; electron energy is given in eV.}
          \label{mu}
   \end{figure}     
       \begin{figure}[h!]
   \centerline{\includegraphics[width=0.95\columnwidth]{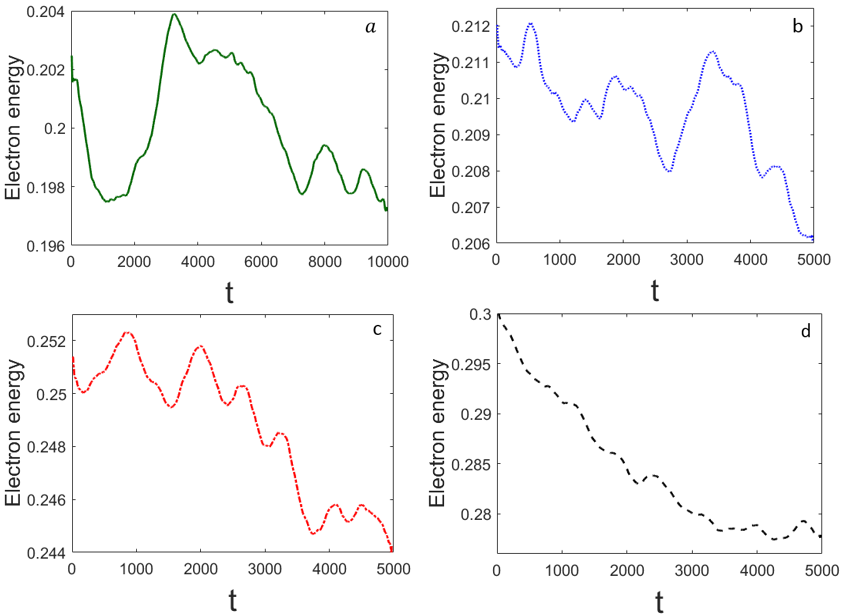}}
   \caption{Decay of the electron energy for a) $\epsilon=0$, b) $\epsilon'=1$ ($\epsilon=6.58$ meV), c) $\epsilon'=5$ ($\epsilon=32.9$ meV), and d) 
   $\epsilon'=10$ ($\epsilon=65.8$ meV); the other parameters are the same of Fig. \ref{amplitude2}, but $\chi'=0.5$ ($\chi=61$ pN). Time $t$ is measured in $10^{-13}$s; electron energy is given in eV.}
          \label{epsilonhalf}
   \end{figure}  
          \begin{figure}[h!]
   \centerline{\includegraphics[width=0.95\columnwidth]{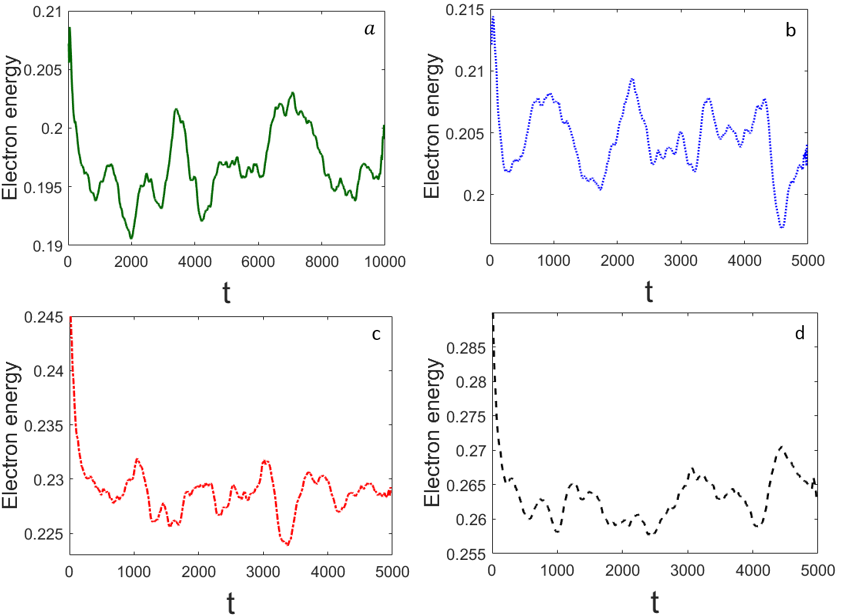}}
   \caption{Decay of the electron energy for a) $\epsilon=0$, b) $\epsilon'=1$ ($\epsilon=6.58$ meV), c) $\epsilon'=5$ ($\epsilon=32.9$ meV), and d) 
   $\epsilon'=10$ ($\epsilon=65.8$ meV); the other parameters are the same of Fig. \ref{amplitude2}, but $\chi'=4$ ($\chi=488$ pN). Time $t$ is measured in $10^{-13}$s; electron energy is given in eV.}
          \label{epsilonfour}
   \end{figure} 
   \clearpage

          \begin{figure}[h!]
   \centerline{\includegraphics[width=0.65\columnwidth]{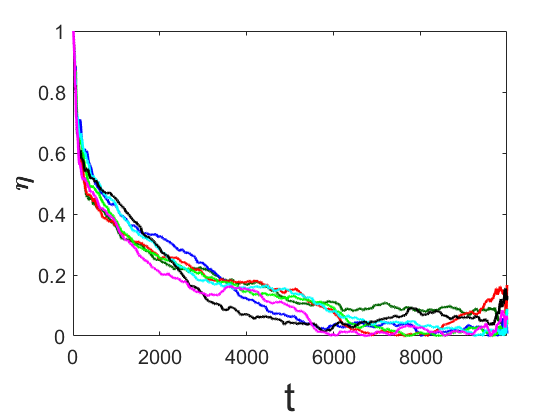}}
   \caption{The spectral entropy $\eta$ is plotted vs time for $\chi'=0.1$ ($\chi=12.2$ pN) (dark green), $\chi'=0.5$ ($\chi=61$ pN) (dark blue), $\chi'=1$ ($\chi=122$ pN) (red), $\chi'=2$ ($\chi=244$ pN )( light green), $\chi'=3$ ($\chi=366$ pN) (light blue), $\chi'=4$ ($\chi=488$ pN) (black), and $\chi'=5$ ($\chi=610$ pN) (purple); the other parameters are the same of Fig. \ref{amplitude2}. Time $t$ is measured in $10^{-13}$s.}
          \label{eta}
   \end{figure}

\section{Concluding remarks}  

The parameter space of the system investigated here is of course very large, thus we have limited our investigation to a basic choice of physically meaningful parameters with respect to the topic that we aimed at better understanding. Then we have checked the robustness of the phenomenology so observed by
changing some parameters, as is the case of the nonlinear coupling constants $\epsilon$ and $\mu$, or the electron-phonon coupling constant $\chi$.
The results actually show that after having given $0.19$eV of initial excitation energy to an electron, the electron wavefunction spreads through the chain by releasing to the phonons only a small fraction of the electron energy, approximately $0.02$eV. This is a somewhat unexpected result but interesting because it helps in understanding why exciting a collective intramolecular oscillation of the BSA protein required a very long time. Of course, the contributions of several fluorophores add up, and the continuous illumination of the labelled proteins with an intense laser light allows to accumulate energy in the protein until the activation threshold of the coherent oscillation of all its atoms is reached and passed over. The phonon part has been simplified with respect to the model derived by the de-quantisation of the original Fr\"ohlich's model \cite{Frohlich1} because the model investigated here has focussed only on the mechanism of down-conversion of the energy of the photons, harvested by the protein through its fluorophores receptors, to the internal vibrations of the chain of amino acids. 
Although no more than $10\%$ of energy is dissipated by electron to phonons, 
it seems that in the studied regime no coherent transport of information can 
occur on the amino acids (as sometimes one could expect in a spin chain model \cite{bayat} )
due to the fact that the electron wave function spreads over all sites.
The model studied here can be easily adapted to estimate the efficiency of other excitation mechanisms of biomolecular collective oscillations, like, possibly, the Coulomb collisions of the phosphate anion produced by the hydrolysis of ATP yielding a momentum transfer on some target electron. Or, as already mentioned in the Introduction, by anisotropic momentum transfer operated by water molecules or small ions resulting in collisional excitation of electrons.
Let us conclude by mentioning that, for a broad class of Hamiltonian systems, long-living Quasi Stationary States (QSS) can be dynamically generated which keep a system out of thermodynamic equilibrium. Among many other systems where QSS are produced \cite{ruffo}, let us mention a beam of fast particles interacting with the set of waves describing a physical system \cite{fanelli,ppcf}, a situation which is reminiscent, for example, of the above mentioned fast phosphate groups - produced by ATP hydrolysis.

\section*{Acknowledgments}
E.F. warmly thanks the Fondazione Cassa di Risparmio di Firenze for having co-funded her PhD fellowship.
M.P. participated in this work within the framework of the project MOLINT which has received funding from the Excellence Initiative 
of Aix-Marseille University - A*Midex,  a French “Investissements d’Avenir” programme. 
R.F. acknowledges support by the QuantERA ERA-NET Co-fund 731473 (Project Q-CLOCKS).
\clearpage
\section*{Appendix}
\begin{center}
\begin{table}[htbp]
\centering

\label{electron-amino-acids}
\begin{tabular}{|c|c|c||c|c|c|}
\hline
          \textbf{Amino acid} &\textbf{EIIP Ry}&\textbf{EIIP eV}& \textbf{Amino acid} &\textbf{EIIP Ry}&\textbf{EIIP eV}
        \\  
        \hline
      Leu&0.0000& 0.0000&     Tyr&0.0516&  0.7017\\
      Ile  &0.0000&  0.0000&   Trp&0.0548&  0.7452\\
      Asn &0.0036& 0.0489&    Gln&0.0761&  1.0349\\
      Gly &0.0050& 0.0680&      Met&0.0823&  1.1192\\
      Val &0.0057& 0.0775&      Ser&0.0829& 1.1274\\
      Glu&0.0058&  0.0788&    Cys&0.0829&  1.1274\\
        Pro&0.0198&  0.2692& Thr&0.0941&  1.2797\\
       His&0.0242&  0.3291&   Phe&0.0946&  1.2865\\
       Lys&0.0371&  0.5045&  Arg&0.0959&  1.3042\\
        Ala&0.0373& 0.5072&   Asp&0.1263& 1.7176\\
   \hline              
\end{tabular}
\caption{Electron-Ion interaction potential (EIIP) value for amino acids. From Ref.\cite{Cosic}.}
\end{table}
\end{center}

\pagebreak [4]

\end{document}